\documentclass[3p,times,authoryear]{elsarticle}

\usepackage{amsmath}

\usepackage{amssymb}

\usepackage{natbib}

\usepackage{graphicx,color}

\usepackage[figuresright]{rotating}

\begin{document}

\begin{frontmatter}




\title{Competition between fast- and slow-diffusing species in
  non-homogeneous environments}

\author[label1]{Simone Pigolotti}
\author[label2]{Roberto Benzi}

\address[label1]{Departament de Fisica, Universitat
  Politecnica de Catalunya Edif. GAIA, Rambla Sant Nebridi 22, 08222
  Terrassa, Barcelona, Spain.}
\address[label2]{Dipartimento di Fisica, Universit\'a
  di Roma ``Tor Vergata'' and INFN, via della Ricerca Scientifica 1,
  00133 Roma, Italy.}

\begin{abstract}
  We study an individual-based model in which two
  spatially-distributed species, characterized by different
  diffusivities, compete for resources. We consider three different
  ecological settings. In the first, diffusing faster has a cost in
  terms of reproduction rate. In the second case, resources are not
  uniformly distributed in space. In the third case, the two species
  are transported by a fluid flow.  In all these cases, at varying the
  parameters, we observe a transition from a regime in which diffusing
  faster confers an effective selective advantage to one in which it constitutes
  a disadvantage. We analytically estimate the magnitude of this
  advantage (or disadvantage) and test it by measuring fixation
  probabilities in simulations of the individual-based model. Our
  results provide a framework to quantify evolutionary pressure
  for increased or decreased dispersal in a given environment.
\end{abstract}


\end{frontmatter}



\section{Introduction}

Biological species have evolved complex mechanisms to move in
space. Examples range from bacterial movement by means of flagella to
the capacity of swimming and flying of higher organisms. Rationalizing
the evolutionary significance of movement is not an easy task, as the
need to move in space can be determined by several needs
\citep{dieckmann1999evolutionary}, such as the search for resources,
the attempt of escaping predation or competition by conspecific, and
the search for mates. On the downside, motility has a 
metabolic cost, which becomes particularly relevant for microorganisms
swimming at low Reynolds number \citep{purcell1977life}. Moreover, in
some circumstances, an increased motility can lead to an increased
predation risk, so that a less conspicuous movement strategy can be
advantageous \citep{visser2009swimming,bianco2014analysis}. Finally,
in the absence of chemotaxis or environmental cues, a strongly motile
species can easily abandon a patch full of resources.  For sessile
species, similar tradeoffs apply to seed-dispersal strategies
\citep{may1977dispersal,comins1980evolutionarily}

The combined presence of these contrasting effects implies that, in a
given ecological setting, it is often difficult to determine whether
evolutionary pressure tends to increase or decrease species
motility. It is therefore not surprising that, on the modeling side,
there exists a fairly vast literature, where different models often
reach contrasting conclusions. For example, an analysis by
\citet{dockery1998evolution}, based on deterministic
reaction-diffusion equations, concludes that it is always advantageous
to adopt a less diffusive strategy. By means of a similar argument,
\citet{hastings1983can} concluded that, in a time-independent
environment, evolutionary stable strategies do not involve dispersal.
However, results from stochastic individual-based models
\citep{kessler2009fluctuations,waddell2010demographic,lin2014demographic,pigolotti2014selective,novak2014habitat,lin2015demographic}
show that diffusing faster can indeed be advantageous. Also in the
context of seed dispersal strategies, the classic analysis by
\citet{may1977dispersal} shows that a certain degree of dispersal is
beneficial also in spatially homogeneous environments, see also
\citet{comins1980evolutionarily}.

In this paper, we show that an ``effective'' selective advantage (or
disadvantage) can be associated to a higher diffusivity in different
ecological settings. The effective selective advantage can be used to
explicitly quantify whether evolutionary pressure promotes or
disfavour an increased diffusivity. To this aim, we study a general
individual-based model in which two different species, or alleles,
compete stochastically in space. The model is similar in spirit to
Kimura's stepping stone model \citep{kimura1964stepping}, except that
we consider a continuous space rather than a discrete array of
island. Individual belonging to the two species diffuse in space with
different diffusivities. Reproduction rates can depend on space and on
the species. As the dynamics of the model is stochastic, the fixation
of one of the two species is a random event. We study which ecological
conditions lead to a bias in the fixation probability towards either
the fast or the slow species and analytically quantify the selective
advantage causing this bias.

In particular, we consider three different settings, which are
representative of common ecological tradeoffs. In the first, the
environment is spatially homogeneous, but the fastest species
reproduces at a slower rate due to the cost of mobility. { This
  simple case is useful to introduce the basic concepts and in
  particular to quantify the selective advantage for fast diffusing
  species due to demographic stochasticity.} In the second setting,
the two species reproduce at equal rate but the environment is
spatially non-homogeneous. In the third case, reproduction rates are
equal, the environment is non-homogeneous, and the two species are
transported by a compressible velocity field. This latter case can be
seen as an idealized example of competition in a marine
environment. In all three cases, we find that, depending on
parameters, diffusing faster can be either advantageous or
disadvantageous, depending on trade-offs between different
ecological forces that we explicitly quantify. We conclude by
discussing our results in the light of existing literature.

\section{Methods}\label{sec:model}

\subsection{Model}\label{sub:mod}

We consider an individual-based model in which two species (or alleles)
$A$ and $B$ compete with each other
\citep{pigolotti2012population,pigolotti2013growth}.  Individuals of
the two species diffuse in a one-dimensional space with different
diffusivities $D+\delta D$ and $D$ respectively, modeling different
spatial motilities. Without loss of generality, we consider the case
in which species $A$ diffuses faster, $\delta D>0$. Species $A$ and
$B$ reproduce stochastically at rates $\mu(x)(1+s)$ and $\mu(x)$
respectively, where $\mu(x)$ represents the density of resources at
spatial coordinate $x$ and $s$ is the reproductive advantage (if
positive) for the fastest species. The death rates of species $A$ and
$B$ depend on the local density of individuals.  In Section
\ref{sub:adv}, we also consider a case in which the species are
transported by a velocity field $v(x)$, for example representing
aquatic currents for marine organisms. { Further details on the
implementation of the individual-based model are discussed in \ref{app:model}.}

\begin{figure}[htb]
\includegraphics[width=9cm]{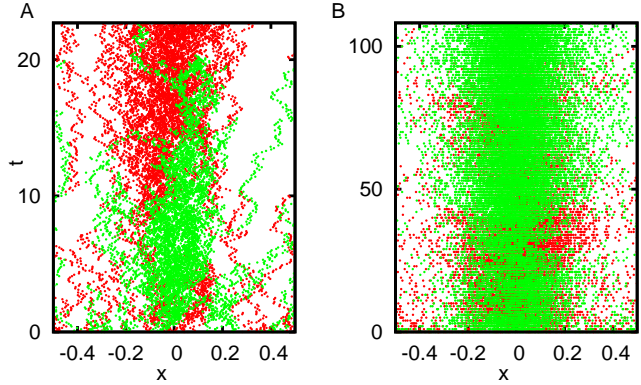}
\caption{Examples of the dynamics of the individual-based model. In
  both panels, $\mu(x)$ is a Gaussian distribution with average zero
  and variance $\sigma=0.1$. The red species is the fastest: in both
  panels, $\delta D/D=0.2$ and $D= 10^{-3}$. The two species are
  reproductively neutral, $s=0$ and are initially present at the same
  density.  The total number density is {\bf A} $N=50$ and {\bf B}
  $N=300$. By performing several realizations of the two simulations,
  we find that the fixation probability of the fastest species in {\bf
    A} is $P_{fix}\approx 0.62$ and in {\bf B} is $P_{fix}\approx 0.57$.
  \label{fig_ill} }
\end{figure}

An example of simulation of the model is shown in
Fig. (\ref{fig_ill}), where the two species compete for a localized
patch of resources. Simulations are run until fixation, i.e. the time
at which either species $A$ or $B$ goes extinct. We anticipate that
all parameters of the model, including the size of the total
population size $N$ as in the case of the figure, can be responsible
for biasing fixation towards the fast or the slow diffusing species.  The
macroscopic dynamics can be analyzed by deriving stochastic evolution
equations for the concentrations $c_A(x,t)$ and $c_B(x,t)$ of the two
species \citep{pigolotti2012population,pigolotti2013growth}, that read

\begin{eqnarray}\label{generalmodel}
\partial_t c_A&=&\overbrace{-\nabla[v(x)c_A]}^{\mathrm{Advection}}+\overbrace{(1+s)\mu(x) c_A}^{\mathrm{
    Growth}}-\overbrace{c_A (c_A+c_B)}^{\mathrm{
    Competition}}+\overbrace{(D+\delta D) \nabla^2 c_A}^{\mathrm{Diffusion}}
+\overbrace{\sigma_A  \xi_B(x,t)}^{\mathrm{Fluctuations}}\nonumber\\
\partial_t c_B&=&-\nabla[v(x)c_B]+\qquad \mu(x)\ \ c_B-c_B (c_A+c_B)
+\quad \ D
\nabla^2 c_B\quad\ +\ \sigma_B  \xi_B(x,t)
\end{eqnarray}
where $\xi_A(x,t)$, $\xi_B(x,t)$ are Gaussian, independent,
delta-correlated noise sources representing demographic stochasticity.
The noise amplitudes are
$\sigma_A^2=c_A[\mu(x)+c_A+c_B]/N$ and similarly for $c_B$.
 The intraspecific and interspecific
competition coefficient in Eqs. (\ref{generalmodel}) are set to
one by an appropriate choice of the density-dependent death rates and
the interaction length in the individual-based model, see
\citet{pigolotti2012population,pigolotti2013growth} and
\ref{app:model} for details.

The dynamics embodied in Eqs. (\ref{generalmodel}) is characterized by
a very rich phenomenology \citep{pigolotti2013growth}. In this broad
framework, we focus on the specific problem of understanding when
having a larger diffusivity $\delta D>0$ confers a selective advantage
or disadvantage to species $A$. We shall study this problem in
different settings, where different terms in Eqs. (\ref{generalmodel})
dominate. A possible way to analytically study the problem is the
following. Consider Eqs. (\ref{generalmodel}) and change variables to
the total concentration $c_T(x,t)=c_A+c_B$ and the relative fraction
of species $A$, $f(x,t)=c_A/c_T$.  We begin by writing the equation
for $c_T$. We always consider cases in which the difference in growth
rate and diffusivities are both small, $s\ll 1$ and $\delta D/D\ll
1$.
Under these approximations, $c_T$ evolves according to a closed
equation

\begin{equation}
\partial_t c_T=-\nabla[v(x)c_T]+\mu(x) c_T- c_T^2 +D \nabla^2 c_T +\delta D \nabla^2 c_A
+\sqrt{\sigma^2_A+\sigma^2_B} \xi .
\end{equation}

When the diffusion length scale is much smaller than the typical
length scale of the gradient of $\mu(x)$ (in other words, $D\ll 1$)
{ and the velocity field vanishes, $v(x)=0$, } the stationary
solution can be approximated as $c_T\approx \mu(x)$, { i.e. the
  total population is close to an ideal free distribution (IFD)
  \citep{fretwell1970}. In the last part of the Results section, we
  also consider an example in which $v(x)\neq 0$ and
  the distribution of the total population is not necessarily close to
  an IFD.
}

Once { the equilibrium value of } $c_T$ is known, it can be
substituted in the equation for the fraction $f$, which is the
relevant quantity to determine which of the two species fixates. By
analyzing this equation, we shall see that one can identify the
effects leading to selective advantages to the fastest or the slower
species.

\subsection{Fixation probabilities}

By repeated simulations of the individual-based model, we can estimate
the fixation probability $P_{fix}$ of the fastest species for a given
setting. In simple evolutionary models such as Wright-Fisher dynamics,
a selective advantage biases the fixation probability according to
Kimura's formula \citep{kimura1962probability}
\begin{equation}\label{kimurafix}
P_{fix}=\frac{1-e^{-Nf_0s}}{1-e^{-Ns}}
\end{equation}
where $f_0$ is the initial fraction of mutants.  { In simple
  cases, the same formula holds when considering a spatially extended
  population
  \citep{maruyama1970effective,whitlock2003fixation,doering2003interacting}}.
Examples are the stepping-stone model \citep{kimura1964stepping} and
similar spatial models that can be described in the continuum limit by
the stochastic Fisher equation

\begin{equation}\label{stoch_fish}
\partial_t f(x,t)= s f(1-f)+D\nabla^2 f +\sqrt{\frac{2f(1-f)}{N}}
\xi(x,t) .
\end{equation}

Eqs. \ref{generalmodel} are more complex than (\ref{stoch_fish}) and
it is not obvious that, in general, Eq. \ref{kimurafix} for the
fixation probability should
hold. Nevertheless, we shall see that, in some cases, the equation for
relative fraction of the fastest species $f$ can be cast in a form
similar to Eq. (\ref{stoch_fish}). This allows to define an
``effective'' selective advantage $s_{eff}$ as the coefficient of the
term proportional to $f(1-f)$ in the resulting equation.

Numerically, the effective selective advantage can be inferred from 
the fixation probability { estimated by simulating several
  realizations of the individual-based model}. In the simple case
of  an equal initial fraction of individuals of the two species,
$f_0=1/2$, Eq. (\ref{kimurafix}) can be analytically inverted,
yielding

\begin{equation}\label{eff_s}
s_{eff}=\left\{ 
\begin{array}{c}
-\frac{2}{N}\log\left(\frac{1-\sqrt{1-4P_{fix}(1-P_{fix})}}{2P_{fix}}\right)\quad
P_{fix}\le \frac{1}{2}\\
-\frac{2}{N}\log\left(\frac{1+\sqrt{1-4P_{fix}(1-P_{fix})}}{2P_{fix}}\right)\quad P_{fix}>\frac{1}{2}
\end{array}
\right. .
\end{equation}

This expression allows for a direct comparison between simulations of
the individual-based models and predictions of our theory. { In
  the second part of the Results section, we will consider one case in
  which the nutrients are not homogeneously distributed in space. In
  this case, we are unaware of mathematical results ensuring 
  that Eq. (\ref{kimurafix}) and consequently Eq. (\ref{eff_s}) hold. We will
  check numerically that this is the case and, in particular, that no
  density-dependent effects appear as the initial fraction of mutants
  $f_0$ is varied.    }

\section{Results}\label{sec:results}

{
\subsection*{The effect of different diffusivities in the homogeneous case}
}
We begin with the simple case in which the nutrient is
spatially homogeneous, $\mu(x)=1$ and advecting flows are absent,
$v=0$. In this case, the fastest species can take advantage of number
fluctuations, as demonstrated in \citep{pigolotti2014selective} and
Appendix A. This subtle effect is purely due to stochasticity and is
absent in the deterministic limit of $N\rightarrow \infty$.

It is interesting to analyze a case in which enhanced motility comes at the expense
of a reduced reproduction rate. This setting represents a common
evolutionary dilemma, where at fixed metabolic budget species have to
decide how much energy to invest into movement respect to
reproduction. In the general system of equations (\ref{generalmodel}),
this correspond to the choice $\delta D>0$ and $s<0$.

\begin{figure}[htb]
\includegraphics[width=9cm]{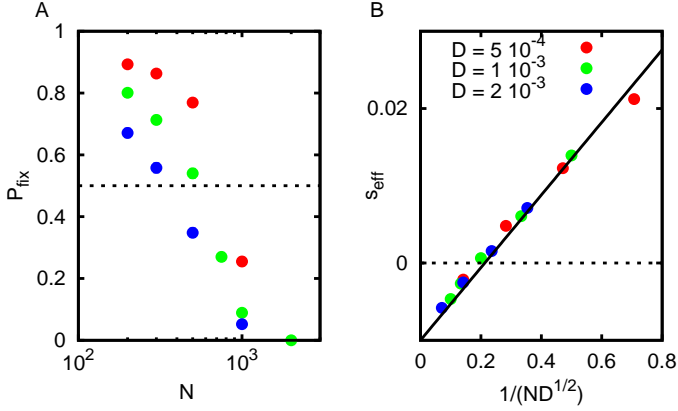}
\caption{ {\bf A} Fixation probabilities at varying $N$ and
  $D$. { In this and the following figures, error bars are either
    explicitly shown or smaller than the size of the symbols.}  {\bf
    B} corresponding rescaling according to Eq. \ref{scal1}. The
  system is spatially homogeneous. Parameters are: $\delta
  D/D=0.2$. The slowest species reproduces at a slower rate,
  $s=-0.01$. In both panels, the dashed lines mark the transition
  between the regions in which diffusing faster is advantageous or
  disadvantageous. { The continuous black line represents a
    linear fit of the data (see Eq. \ref{scal1}) with $\alpha$ as a free parameter, from
    which we obtain $\alpha\approx 0.235$.}\label{fig_tradeoff} }
\end{figure}

We proceed as described in Section \ref{sub:mod}.  The equation for
$f=c_A/c_T$ reads

\begin{equation}\label{new_eq2}
\partial_t f = D \nabla^2 f+\delta D(1-f)\nabla^2 f +
sf(1-f)+\sqrt{\frac{2f(1-f)}{N}}\xi ,
\end{equation}

where we already substituted the stationary value of the total
concentration, $c_T=1$. For $|s|\ll 1$ and $\delta D/D\ll1$, it can be
shown that, on average, the term $\delta D(1-f)\nabla^2 f$ is
proportional to $ \delta D f(1-f)/(ND^{3/2})$, see \ref{app:stochadv} and
\citep{pigolotti2014selective}.  This means that, effectively, the
advantage to the fastest species due to number fluctuations is
proportional to the combination of parameters $\delta D/(ND^{3/2})$.
Substituting this result into (\ref{new_eq2}), one obtains an equation
having the same form as Eq. (\ref{stoch_fish}), with an effective
selective advantage equal to
\begin{equation}\label{scal1}
s_{eff}=s+\alpha\ \delta D /(ND^{3/2})
\end{equation}
where $\alpha$ is a positive constant. { The value of $\alpha$
  can be mathematically estimated using different strategies. In
  \ref{app:stochadv} we summarize the calculation by
  \citet{pigolotti2014selective}, which is based on the continuos
  equation with the genetic drift. Alternatively, one can start from a
  discrete formulation of the same problem as done by
  \citet{novak2014habitat}. In the following, we will check the
  prediction of Eq. (\ref{scal1}) and simply estimate the value of $\alpha$
  numerically.} Eq. (\ref{scal1}) predicts that the effective
selective advantage to the fastest species, derived the from fixation
probability via Eqs. \ref{eff_s}, must increase linearly on the
combination of parameter $1/(ND^{1/2})$.  We performed numerical
simulations of the individual-based model at fixed value of
$\delta D/D$ and $s$ and for different values of $D$ and $N$. We
measured $P_{fix}$, as shown in Figure (\ref{fig_tradeoff}A), and
inferred the effective selective advantage via Eqs. \ref{eff_s}. As
predicted by Eq. (\ref{scal1}), one finds that the effective selective
advantage depends linearly on the combination of parameter
$1/(ND^{1/2})$, see Figure (\ref{fig_tradeoff}B). { A
  linear fit using Eq. (\ref{scal1}) permits to estimate the value of
  $\alpha\approx 0.235$. }

Notice that, at increasing $N$ (or $D$), one finds a transition
between a regime in which it is advantageous to diffuse faster,
because of the larger demographic fluctuations, to one in which it is
not convenient to pay the price of a smaller reproduction rate. Such
transition is marked with a dashed line in both panels of Figure
(\ref{fig_tradeoff}).  { We conclude this section by remarking
  that, in practical cases, $s$ and $\delta D$ are not independent
  variables. In particular, because of energy constraint, the mutants
  with a higher mobility $\delta D$ pay a more severe cost in
  terms of the difference in reproduction rate $s$. Assuming a
  specific functional relation between $s$ and $\delta D$, one can to
  obtain from Eq. (\ref{scal1}) the optimal value of $\delta D$ by
  simply maximizing $s_{eff}$ under the chosen constraint. }

\subsection*{Spatially inhomogeneous case}

We now consider a case in which the growth rates of the two species
are identical, $s=0$, but the nutrients $\mu(x)$ are not uniformly
distributed in space. Also in this case, we start by deriving the
dynamics for the fraction of the fast species $f$
\begin{equation}
\partial_t f = [D+(1-f) \delta D] \left[\nabla^2 f+\frac{2(\nabla f)(\nabla
    c_T)}{c_T}\right]+\delta D f(1-f)\frac{\nabla^2 c_T}{c_T}+\mathrm{noise}
\end{equation}
where, for small $D$,   $c_T$ can be replaced by its stationary value
$\mu(x)$ as previously discussed. The
noise amplitude  is $\sigma^2_f=f(1-f)[\mu(x)+c_T]/(Nc_T)\approx
2f(1-f)/N$. Substituting, we obtain 
\begin{equation}\label{main_eq}
\partial_t f = D \left[\nabla^2 f+\frac{2(\nabla f)(\nabla
    \mu(x))}{\mu(x)}\right]+\delta D\left[(1-f)\nabla^2 f+
  f(1-f)\frac{\nabla^2
    \mu(x)}{\mu(x)}\right]+\sqrt{\frac{2f(1-f)}{N}}\xi .
\end{equation}

Eq. (\ref{main_eq}) is too complex to allow any analytic study. However, it is possible to understand
the basic feature of the system dynamics by a reasonable interpretation of  different terms in
Eq. (\ref{main_eq}).  The two terms in the first square brackets
represent diffusion plus an effective advection term. The latter can
be understood thinking that the relative concentration $f$ tends to be
transported from regions of higher total density to regions of lower
one. Importantly, both terms are symmetric in $f \leftrightarrow
(1-f)$ and are not function of the diffusivity difference $\delta D$. Thus,
they can not be directly responsible for biasing fixation
probabilities towards one of the two species.

Conversely, terms in the second square brackets are proportional to
the diffusivity difference $\delta D$ and thus break the symmetry
between the two species. The first term is identical to that discussed
in the previous section, providing a stochastic advantage to the
fastest species. We assume that, also in this non-homogenous setting,
its average is proportional to
$\langle (1-f)\nabla^2 f\rangle\propto f(1-f)/(ND^{3/2})$ as in the
homogeneous case, where $\langle \dots \rangle$ denotes a spatial
average. The second term does not depend neither on the diffusion nor
on $N$. In a non-homogenous situation, the total population tends to
concentrate around maxima of $\mu(x)$, where $\nabla^2 \mu(x)<0$.
This argument suggests that the average of this term over the
distribution of the total population must be negative. This can be
interpreted as the advantage due to the spatial inhomogeneity for the
less agile species.

\begin{figure}[htb]
\includegraphics[width=9cm]{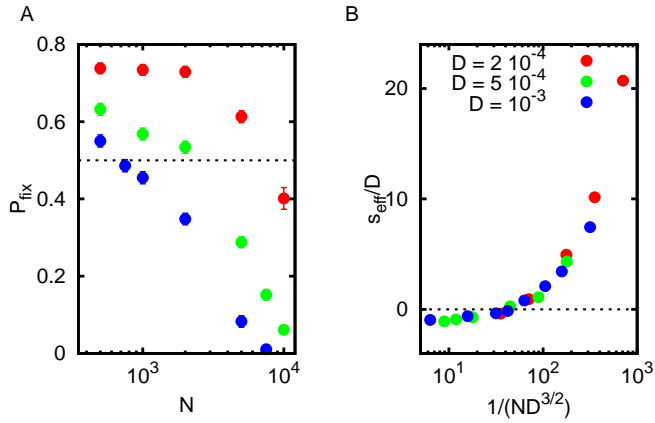}
\caption{{\bf A} Fixation probabilities in a spatially nonhomogenous
  system at varying $N$ and $D$ and {\bf B}
  corresponding rescaling according to Eq. \ref{seff2}.  The growth
  rate depends on space as
  $\mu(x)=e^{-x^2/(2\sigma^2)/}\sqrt{2\pi \sigma^2}$.
Other parameters are: $\delta D/D=0.2$,  $\sigma=0.2$ and $s=0$. The dashed line in
panel {\bf A} marks the transition between the regions in which diffusing
  faster is advantageous or not.
  \label{fig_inhom} }
\end{figure}

If the above assumptions hold, the effective selective advantage in
this case is equal to
\begin{equation}\label{seff2}
s_{eff}=\delta D\left[ \frac{\alpha}{ND^{3/2}}+ K\right] .
\end{equation}
where the first term in the square brackets is the noise-induced
advantage discussed in the previous section, and $K<0$ is a constant
related to the average value of $\nabla^2 \mu(x)/\mu(x)$, { as
  later discussed.} As before, the balance between these two effects
determines whether the fastest or the slowest species fixates with
higher probability. Fixation probabilities for this model at varying
$N$ and $D$ are shown in Fig. (\ref{fig_inhom}A). Also in this case, a
one can observe a transition between regimes in which diffusing faster
is advantageous or not.  Equation \ref{seff2} predicts that,
performing simulations at fixed relative diffusivity difference,
i.e. $\delta D/D=\mathrm{const}$, the quantity $s_{eff}/D$ should be a
function of the combination of parameters $1/(ND^{3/2})$ only. This
prediction is very well verified in the simulations, as shown in
Fig. (\ref{fig_inhom}B).

\begin{figure}[htb]
\includegraphics[width=9cm]{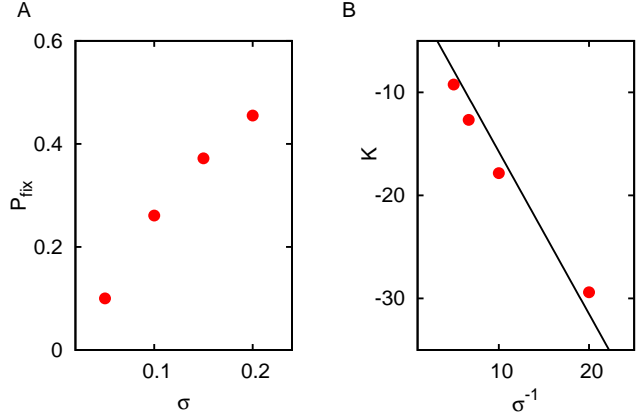}
\caption{{\bf A} Fixation probabilities in a spatially nonhomogenous system at
  varying the characteristic size of the resource patch $\sigma$ and
  corresponding effective selective advantage. Parameters are: $D=10^{-3}$, $N=10^3$ and
  $\delta D/D=0.2$. { {\bf B} Plot of $K=(s_{eff}/\delta D)-\alpha/(ND^{3/2})$
  versus $\sigma^{-1}$ from the same data. The black line is a linear fit 
  \label{fig_inhom2} according to the estimate $K = -\beta\
  \sigma^{-1}$, from which we obtain $\beta\approx 1.6$.} }
\end{figure}

{ Let us study in more details the dependence of the constant $K$ on
  the characteristic size of the resource patch $\sigma$ { for
    the case in which $\mu(x)$ is Gaussian,
    $\mu(x)=(2\pi\sigma^2)^{-1/2}\exp(-x^2/2\sigma^2)$
  }. Fig. \ref{fig_inhom2}A shows that, indeed, the fixation
  probability of the fastest species increases as $\sigma$ is
  increased, implying that $K$ must increase (i.e. decrease in
  absolute value) with $\sigma$. To be more quantitative, let us focus
  on the term proportional to $\nabla^2 \mu(x)/\mu(x)$ in
  Eq. \ref{main_eq}

\begin{equation}
\delta D\left\langle
  f(1-f)\frac{\nabla^2
    \mu(x)}{\mu(x)}\right\rangle=\delta D\int\ dx\ f(1-f)\ 
\frac{x^2-\sigma^2}{\sigma^4}=\frac{\delta D}{\sigma}\int \ dy\  f(1-f)
(y^2-1)
\end{equation}
where $y=x/\sigma$.  This calculation suggest that, at the leading
order, $K$ can be estimated as $K = -\beta\ \sigma^{-1}$,
where $\beta$ is a constant order $1$.  This prediction is verified in
Fig.  \ref{fig_inhom2}B.  }

\begin{figure}[htb]
\includegraphics[width=9cm]{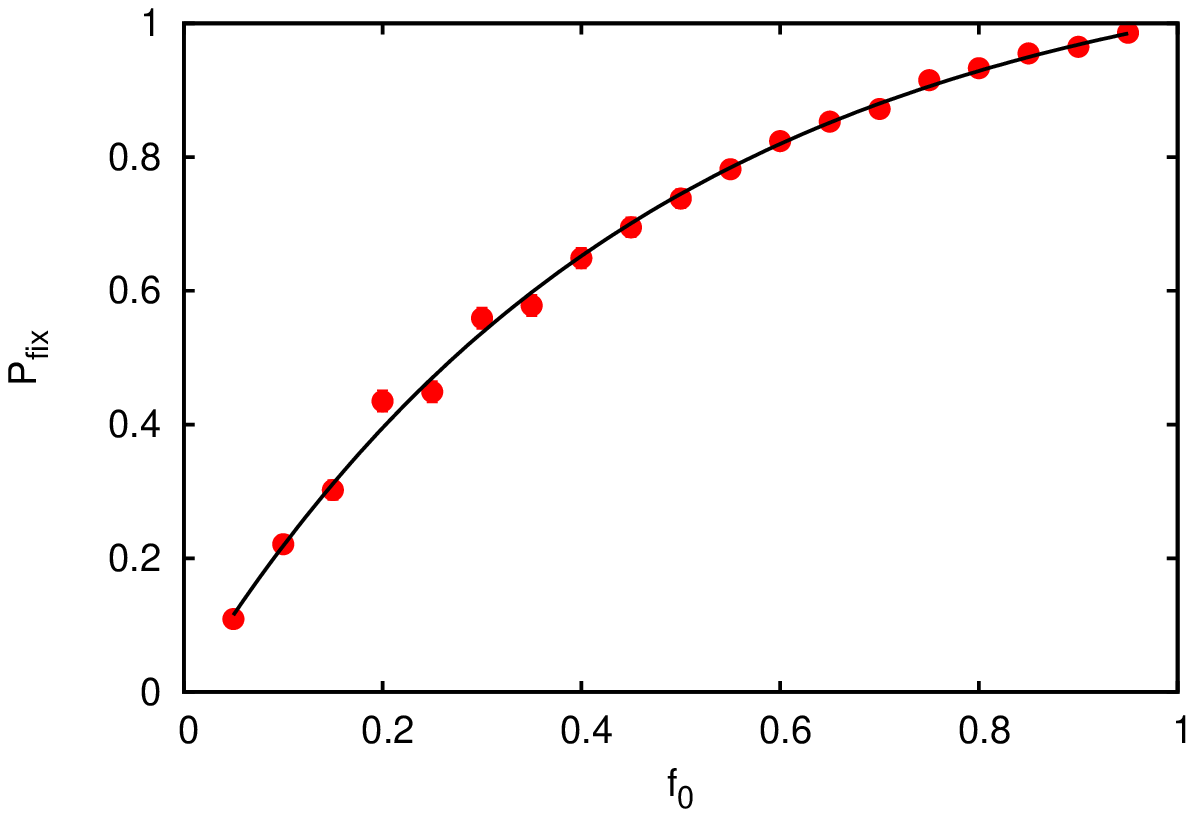}
\caption{{ Fixation probabilities in a spatially nonhomogenous system as
  a function of the initial fraction of mutants $f_0$. For all values
  of $f_0$, individuals of $A$ and $B$ are uniformly distributed in
  space at $t=0$. The continuous
  line is the prediction of Kimura's formula, Eq. \ref{kimurafix}. The
  effective selective advantage $s_{eff}\approx 0.0043$ has been
  obtained via a fitting procedure. Parameters are: $D=2\ 10^{-4}$,
  $N=500$, $\sigma=0.2$ and $\delta D/D=0.2$. }\label{fig_densitydep} }
\end{figure}

  We remark that, in a spatially inhomogeneous system, the fixation
  formula of Eq. \ref{kimurafix} is not guaranteed to be valid as
  non-homogeneities could potentially lead to a density-dependent
  selective advantage. To check this possibility, we studied the
  dependence of the fixation probability as a function of the initial
  fraction of mutants.  The results in Fig. \ref{fig_densitydep} shows
  that, choosing spatially homogeneous initial conditions but at
  different initial fraction $f_0$, fixation probabilities are very
  well described by Kimura's formula also in this scenario. This confirms
  that density-dependent effects are absent or unimportant.  

\subsection*{Advection}\label{sub:adv}

In this section, we study the combined effect of a non-uniform growth rate
$\mu(x) =\exp[-x^2/(2\sigma^2)]$ and a velocity field $v(x)=-\gamma x$
attracting individuals toward the point $x=0$. 

{ This velocity field can be thought of as a simple example of
  the behavior close to sinks of a turbulent velocity field
  \citep{pigolotti2012population}. Indeed, although three-dimensional
  velocity fields in the ocean are obviously incompressible,
  additional forces such as gyrotaxis on small scales
  \cite{durham2013turbulence} or upwelling/downwelling on larger
  scales \cite{benzi2012population} can be taken into account by
  considering in an ``effective'' compressible velocity field
  experienced by microorganisms.  }

For simplicity, we start by analyzing the problem in the absence of
demographic fluctuations, $N\rightarrow \infty$.  Proceeding as
before, we obtain an equation for $f$:
\begin{equation}
\label{eqf}
\partial_t + v(x) \nabla f = (D+\delta D) \Delta f + 2(D+\delta
D) [\nabla \log(c_T)] \nabla f 
+ \delta D \Delta f(1-f)+ \frac{\Delta c_T}{c_T} f(1-f) .
\end{equation}

It is useful to notice that both $\mu(x)$ and $v(x)$ effectively
concentrate the total population $c_T$ in a limited region of space.
In particular, for small values of $D$ and in the absence of the
velocity field, we argued that the non-homogeneous nutrient
distribution would lead to a total population density $c_T \approx
\mu(x)$.  The velocity field alone would concentrate $c_T$ in a
similar way, leading to $c_T \propto \exp[-\gamma x^2/(2D)]$.

However, the characteristic spatial scale at which the population is
concentrated by the two effects is different: $\sigma$ for
non-homogenous $\mu(x)$ and $l \equiv \sqrt{D/\gamma}$ for the
velocity field. It can be expected that the effect of $\delta D$ would
depend on the adimensional ratio $ r \equiv l /\sigma$ of these two different
scales, controlling which of these two effects dominates. For $r\ll1$,
the velocity field controls the width of the total population. In this
case, the species diffusing faster has a selective advantage, as it
can invade more easily upstream regions and therefore invade the
system \citep{pigolotti2012population}.  For $r\gg 1$, the dominant
effect is given by the non-homogeneous nutrient distribution. In this
case, the faster species has a selective disadvantage, for the
argument presented in previous section. An analytical argument
obtained by expanding the function $f$ around the point $x=0$ support
the fact that the transition should occur exactly at $r=1$, see
Appendix B.

\begin{figure}[h]
  \begin{center}
    \includegraphics[width=0.50\textwidth]{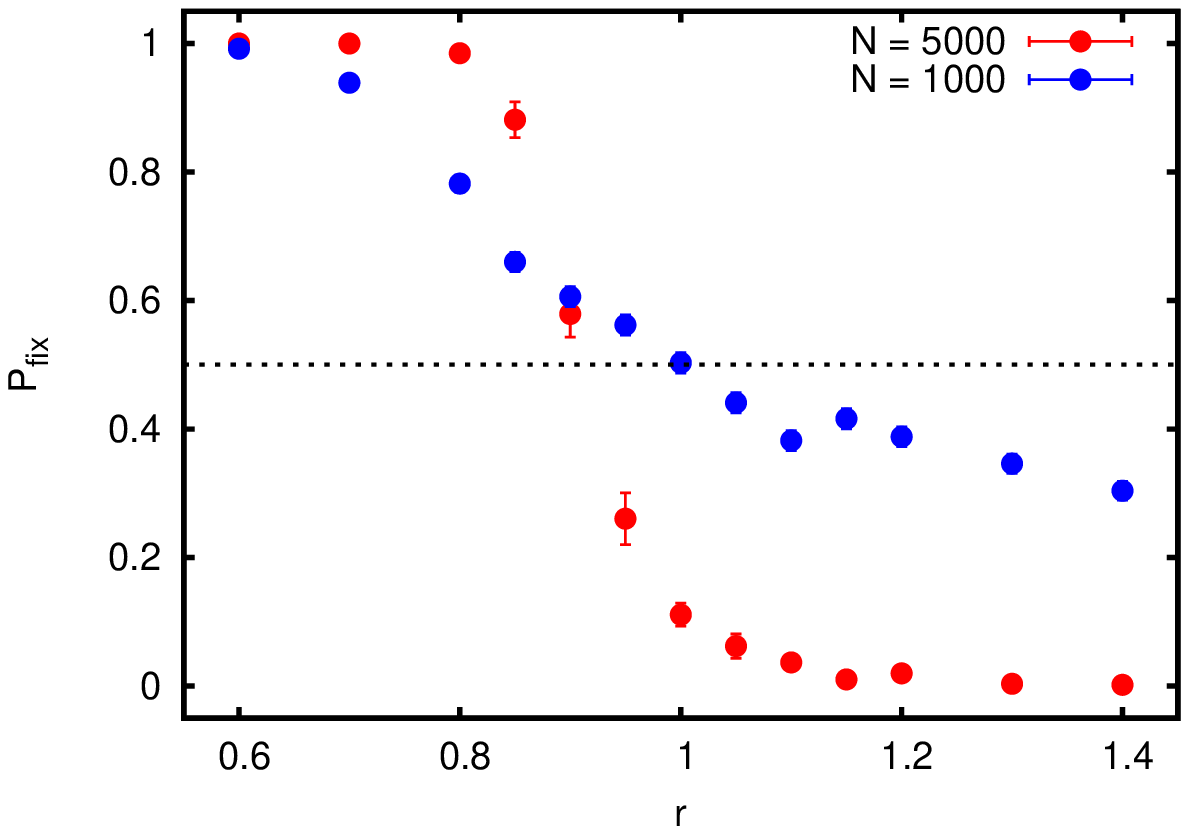}
  \end{center}
  \caption{Dynamics in the presence of an advective linear
    field. Parameters are $D=10^{-3}$, $\delta D/D=0.2$,
    $\sigma=0.1$. The ratio of length scales $r$ is varied by tuning
    $\gamma$.The dashed line marks the transition between a selective
    advantage and a disadvantage to the fastest species.}
  \label{figflo2}
\end{figure}

The transition between these two regimes can be observed in the
numerical simulations of figure (\ref{figflo2}). For large population
size (red dots), one observe a behavior very close to the theoretical
prediction, although the transition seems to occur at a value of $r$
slightly smaller than one. For smaller population size (blue dots) the
transition is more smoothed, as noise plays a more important role in
this case. Notice also that, for large $r$, the fixation probability
of the fast species is significantly larger than zero. This is due to
the previously discussed stochastic advantage for the fast species
which also acts in this case.

{ Following \citet{novak2014habitat}, the transition at $r=1$ can
  also be interpreted by considering which of the two populations is
  closer to an IFD.  In our notation, a population characterized
  by a diffusivity $D$ is at an IFD under the condition of balanced
  dispersal
  \citep{doncaster1997balanced,cantrell2010evolution,novak2014habitat}
\begin{equation}
D\nabla^2  \mu(x) + \nabla [v(x)\mu(x)] = 0.
\end{equation}
which is satisfied only when $\gamma=1/\sigma^2$ and $r=1$. For $r<1$,
the population with higher value of $D$ is closer to an IFD and
therefore has a selective advantage. Conversely, for $r>1$ an increase
of diffusivity leads to a larger deviation from an IFD, therefore to a
decrease in the fixation probability.  }

\section{Discussion}\label{sec:discussion}

In this paper, we studied competition between two populations having
different diffusivities. We have shown how the ecological forces
acting on the populations, including non-homogeneities in the resource
distributions and transport by fluid flows, determine the best
competitor in a non-trivial way. All these ingredients can be
systematically analyzed by means of macroscopic stochastic equations
describing the evolution of the two populations in space and
time. This approach allows for explicitly estimate the effective
selective advantage, or disadvantage, granted by diffusing faster in
a given ecological settings.

The three examples discussed in this paper present different
tradeoffs between two ecological forces, one promoting diffusion and
one contrasting it.  Our first example was contrasting demographic
stochasticity, giving an advantage to the fastest species, with a
slower reproduction rate. In the second case, we contrasted
demographic stochasticity with a non-homogenous nutrient distribution,
where the latter confers an advantage to the { slowest } species. We
concluded with a case in which non-homogeneity in the nutrients is
contrasted with a fluid flow concentrating individuals around a
velocity sink. In this latter case, diffusing faster constitutes an
advantage as faster individuals can colonize more easily upstream
regions, from which they can invade.

{ These three examples fit into a more general theoretical
  scheme. In the absence of demographic stochasticity, or other
  temporal fluctuations, evolution tends to advantage the level of
  dispersal that brings the population close to an IFD
  \citet{novak2014habitat}. In the absence of fluid transport, this
  implies that the species diffusing less is always more fit
  \citep{hastings1983can,dockery1998evolution}. However, this is not
  necessarily the case in the presence of intense fluid flows, where
  the {\em fastest} species is closer to an IFD. } Further, taking
into account demographic stochasticity does not only make the outcome
of competition more uncertain, as in classic examples of competing
species having different reproduction rates
\citep{kimura1962probability}; it also confers an additional advantage to the
species diffusing faster. This observation is consistent with previous
studies
\citep{kessler2009fluctuations,waddell2010demographic,lin2014demographic,pigolotti2014selective,lin2015demographic}.
This effect is already appreciable for differences in diffusivities of
a few percent. While we focused for simplicity on one spatial
dimension, analytical and numerical studies show that the effect
becomes even stronger as spatial dimension is increased
\citep{pigolotti2014selective}. Overall, these results challenge the
view that in time-independent environments it is always convenient to
diffuse less \citep{hastings1983can,dockery1998evolution} and suggest
that deterministic models can miss a crucial ingredient to determine
the best dispersal strategy.

It is tempting to interpret the stochastic advantage to fast species
discussed here to the classic argument by \citet{may1977dispersal},
according to which the advantage of diffusing faster is related to
avoidance of competition with conspecific. Indeed, demographic
stochasticity is one of the main ingredients of Hamilton and May's
seed dispersal model. It should however be noted that, in the context
of the model presented in this paper, the intensity of intraspecific
competition is the same for both species because of symmetry. After
neglecting fluctuations of the total density, interspecific
competition depends only on the relative fraction of the two species
and not on their diffusivity. It is therefore not obvious whether one
can interpret the stochastic advantage of diffusing faster in our
model in terms of a reduced interspecific competition.

Other effects not discussed in this paper can be understood by means
of similar arguments. For example, in population models where a
pattern-forming transition organizes species in spatial clusters,
there is a tendency of slower species to be the best competitors
\citep{hernandez2004clustering,heinsalu2013clustering,hernandez2014spatial}. This
phenomenon is resemblant to the case of non-homogeneous nutrients. In
both cases, a non-homogeneous carrying capacity confers an advantage
to less motile species, although in this case the non-homonogeneity is
not explicit but rather due to an instability of the homogeneous
state.

We thank C. Doering and E. Hernandez-Garcia for useful discussions. SP
acknowledges partial support from Spanish Ministry of Economy and Competitiveness and FEDER (project FIS2012-37655-C02-01).

\appendix
{
\section{Individual-based model}\label{app:model}

In this Appendix, we discuss details of the implementation of the
individual-based model.
Individual belonging to the two species are treated as point-like
particles in a one-dimensional space. The coordinate $x$ of
of any given particle evolves according to the Langevin dynamics
\begin{equation}\label{lang_part}
\dot{x}=v(x)+\sqrt{2D_{A,B}}\ \xi(t)
\end{equation}
where $v(x)$ is a velocity field, different from zero only in the case
of Section (\ref{sub:adv}). $D_{A,B}$ is the diffusivity which
depends on the species ($A$ or $B$) and $\xi(t)$ is a white noise. 

In addition, particles can be created (birth event) or removed (death
event) stochastically. Each individual of species $A$ reproduces at
rate $\mu(x)$, while each individual of species $B$ reproduces at rate
$\mu(x)(1+s)$ where $s$ is a selective advantage. Newborn individuals
are placed at the same coordinate as their mother. Defining the
interaction distance $\delta$ eath event occur at a rate
$\tilde{n}_A+\tilde{n}_B$, where $\tilde{n}_{A,B}$ are the number of
individuals found at a distance less than $\delta$ from each
individual. Dead individuals are simply removed from the system.  See
\citep{perlekar2011particle} for further details on the numerical
implementation.

It can be shown that the parameter $\delta$ can be used to tune the
average number of individuals. In particular, defining $\delta=1/N$
and $c_{A,B}(x)=n_{A,B}(x)/N$ where $=n_{A,B}(x)$ is the number
density of individuals, one can derive from the individual-based model
the macroscopic Eqs. (\ref{generalmodel}), see
\citet{pigolotti2012population,pigolotti2013growth}.
}
\section{Stochastic advantage of diffusing faster in homogeneous
  environments}\label{app:stochadv}

In this Appendix, we briefly recall the main result of
\citet{pigolotti2014selective}. Our starting point is the
one-dimensional stochastic differential equation
\begin{equation}
\label{simple}
\partial_t f = D \nabla^2 f + \delta D (1-f)\nabla^2 f  + \sigma \xi
\end{equation}
with $\sigma= \sqrt{ 2\mu f(1-f)/N}$. Calling $\langle\dots\rangle$
the average over space and noise and averaging 
Eq. (\ref{simple}) leads to
\begin{equation}\label{simple2}
\frac{d\ \langle f\rangle}{dt}=\langle (\nabla f)^2\rangle .
\end{equation} 
Since the right hand side of Eq. (\ref{simple2}) is always
non-negative, one can already conclude that the average effect of
having a larger diffusivity is advantageous. We can rewrite
Eq. (\ref{simple2}) in terms of the heterozygosity function,
$H(d,t)=\langle f(x)[1-f(x+d)]+f(x+d)[1-f(x)]\rangle$. After some
simple algebra we get
\begin{equation}\label{simple3}
\frac{d\ \langle f\rangle}{dt}=\frac{\delta D}{2}
\frac{\partial^2}{\partial d^2}H (d,t) |_{d=0} .
\end{equation}
Eq. (\ref{simple3}) is an exact relation between the average growth of
the density of the fastest species and the heterozygosity. To make
progress, we assume $\delta D/D\ll 1$. In this limit, we can estimate
the right hand side of Eq. (\ref{simple3}) at first order in
perturbation theory, i.e. by approximating the heterozygosity with
that calculated in the purely neutral case of $\delta D=0$, which is
explicitly known, see e.g. \citet{korolev2010genetic}. Under this approximation,
we finally obtain
\begin{equation}
\frac{dF(t)}{ dt} =  \frac{ \delta D}{4D\sqrt{\pi \epsilon t_f} } H(0,0)  {G(t/t_f)}
\label{integral3}
\end{equation}
where $G(x)=\exp(x)\mathrm{erfc}(\sqrt{x})$ and $t_f=2DN^2$. The parameter
$\epsilon$ is a constant with the dimension of a time and on the order
of the smallest time-scale of the system, i.e. the generation time
$\epsilon \approx \mu^{-1}$. Proceeding in the same way for the
stochastic Fisher equation, one would obtain $dF/dt= s H(0,0)
G(t/t_f)/2$. By comparing this expression and Eq. \ref{integral3}, we
can finally define an effective selective advantage 
\begin{equation}
s_{eff}=\frac{\alpha\delta D}{ND^{3/2}}
\end{equation}
where we defined the constant $\alpha=(2\sqrt{2\pi\epsilon})^{-1}$.

\section{Transition in a velocity sink}

In this section, we present an approximate theory for the results
of Fig. (\ref{figflo2}).  We assume that the important
region to study is near $x=0$, where the relative density of the fast
species can be approximated as $f(x)\approx f_0+f_1x^2$. We
substitute this assumption in (\ref{eqf}) and disregard all terms
$o(x^4)$. We also assume $c_T=\mu(x)$, which is a good approximation
near the point $r=1$. Equating terms of the same order in $x$ leads to the
system of differential equations
\begin{eqnarray}
\label{1}
{\dot f_0} &=& 2Df_1 + 2 \delta D f_1(1-f_0) - \delta D f_0(1-f_0)   \\
\label{2}
{\dot f_1} &=&   2\Gamma f_1- (4D+5\delta D)\frac{f_1}{\sigma^2} -2
\delta D f_1^2 + 
\frac{\delta D}{\sigma^2} f_1f_0 + \frac{\delta D}{\sigma^4}f_0(1-f_0)
\end{eqnarray}
This system admits two fixed points, $(f_0,f_1)$, equal to
$(0,0)$ and $(1,0)$, corresponding to fixation of one of the two
species respectively. Let us study the stability of the point $0,0$. The
eigenvalues $\lambda_i$ of the linear stability analysis are
\begin{eqnarray}
\lambda_1 &=& \frac{1}{\sigma^2}  \left[ 2D(\xi-2)+ 6 \delta D +
  \delta D \frac{1-\xi}{2-\xi} \right] \\
\lambda_2 &=& -\frac{\delta D}{\sigma^2} \frac{1-\xi}{2-\xi}
\end{eqnarray}
where  $\xi = \gamma \sigma^2/D=r^{-2}$. While
$\lambda_1<0$ for small $\delta D/D$ and $\xi \approx 1$, $\lambda_2$ becomes
positive at $\xi > 1$, where the stable fixed point becomes $1,0$. 
This analysis explains the transition observed close to $r=1$ in figure
(\ref{figflo2}).




\bibliographystyle{elsarticle-harv}
\bibliography{twodiff}

\begin{thebibliography}{31}
\expandafter\ifx\csname natexlab\endcsname\relax\def\natexlab#1{#1}\fi
\expandafter\ifx\csname url\endcsname\relax
  \def\url#1{\texttt{#1}}\fi
\expandafter\ifx\csname urlprefix\endcsname\relax\def\urlprefix{URL }\fi

\bibitem[{Benzi et~al.(2012)Benzi, Jensen, Nelson, Perlekar, Pigolotti, and
  Toschi}]{benzi2012population}
Benzi, R., Jensen, M.~H., Nelson, D.~R., Perlekar, P., Pigolotti, S., Toschi,
  F., 2012. Population dynamics in compressible flows. The European Physical
  Journal Special Topics 204~(1), 57--73.

\bibitem[{Bianco et~al.(2014)Bianco, Mariani, Visser, Mazzocchi, and
  Pigolotti}]{bianco2014analysis}
Bianco, G., Mariani, P., Visser, A.~W., Mazzocchi, M.~G., Pigolotti, S., 2014.
  Analysis of self-overlap reveals trade-offs in plankton swimming
  trajectories. Journal of The Royal Society Interface 11~(96), 20140164.

\bibitem[{Cantrell et~al.(2010)Cantrell, Cosner, and
  Lou}]{cantrell2010evolution}
Cantrell, R.~S., Cosner, C., Lou, Y., 2010. Evolution of dispersal and the
  ideal free distribution. Mathematical biosciences and engineering: MBE 7~(1),
  17--36.

\bibitem[{Comins et~al.(1980)Comins, Hamilton, and
  May}]{comins1980evolutionarily}
Comins, H.~N., Hamilton, W.~D., May, R.~M., 1980. Evolutionarily stable
  dispersal strategies. Journal of theoretical Biology 82~(2), 205--230.

\bibitem[{Dieckmann et~al.(1999)Dieckmann, O'Hara, and
  Weisser}]{dieckmann1999evolutionary}
Dieckmann, U., O'Hara, B., Weisser, W., 1999. The evolutionary ecology of
  dispersal. Trends in Ecology \& Evolution 14~(3), 88--90.

\bibitem[{Dockery et~al.(1998)Dockery, Hutson, Mischaikow, and
  Pernarowski}]{dockery1998evolution}
Dockery, J., Hutson, V., Mischaikow, K., Pernarowski, M., 1998. The evolution
  of slow dispersal rates: a reaction diffusion model. Journal of Mathematical
  Biology 37~(1), 61--83.

\bibitem[{Doering et~al.(2003)Doering, Mueller, and
  Smereka}]{doering2003interacting}
Doering, C.~R., Mueller, C., Smereka, P., 2003. Interacting particles, the
  stochastic fisher--kolmogorov--petrovsky--piscounov equation, and duality.
  Physica A: Statistical Mechanics and its Applications 325~(1), 243--259.

\bibitem[{Doncaster et~al.(1997)Doncaster, Clobert, Doligez, Danchin, and
  Gustafsson}]{doncaster1997balanced}
Doncaster, C.~P., Clobert, J., Doligez, B., Danchin, E., Gustafsson, L., 1997.
  Balanced dispersal between spatially varying local populations: an
  alternative to the source-sink model. The American Naturalist 150~(4),
  425--445.

\bibitem[{Durham et~al.(2013)Durham, Climent, Barry, De~Lillo, Boffetta,
  Cencini, and Stocker}]{durham2013turbulence}
Durham, W.~M., Climent, E., Barry, M., De~Lillo, F., Boffetta, G., Cencini, M.,
  Stocker, R., 2013. Turbulence drives microscale patches of motile
  phytoplankton. Nature communications 4.

\bibitem[{Fretwell and Lucas(1969)}]{fretwell1970}
Fretwell, S.~D., Lucas, H.~R., 1969. On territorial behavior and other factors
  influencing habitat distribution of birds. Acta Biotheoretica 19, 16--36.

\bibitem[{Hamilton and May(1977)}]{may1977dispersal}
Hamilton, W.~D., May, R.~M., 1977. Dispersal in stable habitats. Nature
  269~(5629), 578--581.

\bibitem[{Hastings(1983)}]{hastings1983can}
Hastings, A., 1983. Can spatial variation alone lead to selection for
  dispersal? Theoretical Population Biology 24~(3), 244--251.

\bibitem[{Heinsalu et~al.(2013)Heinsalu, Hern{\'a}ndez-Garcia, and
  L{\'o}pez}]{heinsalu2013clustering}
Heinsalu, E., Hern{\'a}ndez-Garcia, E., L{\'o}pez, C., 2013. Clustering
  determines who survives for competing brownian and l{\'e}vy walkers. Physical
  review letters 110~(25), 258101.

\bibitem[{Hern{\'a}ndez-Garc{\'\i}a et~al.(2014)Hern{\'a}ndez-Garc{\'\i}a,
  Heinsalu, and Lopez}]{hernandez2014spatial}
Hern{\'a}ndez-Garc{\'\i}a, E., Heinsalu, E., Lopez, C., 2014. Spatial patterns
  of competing random walkers. Ecological Complexity.

\bibitem[{Hern{\'a}ndez-Garc{\'\i}a and
  L{\'o}pez(2004)}]{hernandez2004clustering}
Hern{\'a}ndez-Garc{\'\i}a, E., L{\'o}pez, C., 2004. Clustering, advection, and
  patterns in a model of population dynamics with neighborhood-dependent rates.
  Physical Review E 70~(1), 016216.

\bibitem[{Kessler and Sander(2009)}]{kessler2009fluctuations}
Kessler, D.~A., Sander, L.~M., 2009. Fluctuations and dispersal rates in
  population dynamics. Physical Review E 80~(4), 041907.

\bibitem[{Kimura(1962)}]{kimura1962probability}
Kimura, M., 1962. On the probability of fixation of mutant genes in a
  population. Genetics 47~(6), 713.

\bibitem[{Kimura and Weiss(1964)}]{kimura1964stepping}
Kimura, M., Weiss, G.~H., 1964. The stepping stone model of population
  structure and the decrease of genetic correlation with distance. Genetics
  49~(4), 561.

\bibitem[{Korolev et~al.(2010)Korolev, Avlund, Hallatschek, and
  Nelson}]{korolev2010genetic}
Korolev, K., Avlund, M., Hallatschek, O., Nelson, D.~R., 2010. Genetic demixing
  and evolution in linear stepping stone models. Reviews of modern physics
  82~(2), 1691.

\bibitem[{Lin et~al.(2014)Lin, Kim, and Doering}]{lin2014demographic}
Lin, Y.~T., Kim, H., Doering, C.~R., 2014. Demographic stochasticity and
  evolution of dispersion i. spatially homogeneous environments. Journal of
  mathematical biology, 1--32.

\bibitem[{Lin et~al.(2015)Lin, Kim, and Doering}]{lin2015demographic}
Lin, Y.~T., Kim, H., Doering, C.~R., 2015. Demographic stochasticity and
  evolution of dispersion ii: Spatially inhomogeneous environments. Journal of
  mathematical biology 70~(3), 679--707.

\bibitem[{Maruyama(1970)}]{maruyama1970effective}
Maruyama, T., 1970. Effective number of alleles in a subdivided population.
  Theoretical population biology 1~(3), 273--306.

\bibitem[{Novak(2014)}]{novak2014habitat}
Novak, S., 2014. Habitat heterogeneities versus spatial type frequency
  variances as driving forces of dispersal evolution. Ecology and evolution
  4~(24), 4589--4597.

\bibitem[{Perlekar et~al.(2011)Perlekar, Benzi, Pigolotti, and
  Toschi}]{perlekar2011particle}
Perlekar, P., Benzi, R., Pigolotti, S., Toschi, F., 2011. Particle algorithms
  for population dynamics in flows. In: Journal of Physics: Conference Series.
  Vol. 333. IOP Publishing, p. 012013.

\bibitem[{Pigolotti and Benzi(2014)}]{pigolotti2014selective}
Pigolotti, S., Benzi, R., 2014. Selective advantage of diffusing faster.
  Physical review letters 112~(18), 188102.

\bibitem[{Pigolotti et~al.(2012)Pigolotti, Benzi, Jensen, and
  Nelson}]{pigolotti2012population}
Pigolotti, S., Benzi, R., Jensen, M.~H., Nelson, D.~R., 2012. Population
  genetics in compressible flows. Physical review letters 108~(12), 128102.

\bibitem[{Pigolotti et~al.(2013)Pigolotti, Benzi, Perlekar, Jensen, Toschi, and
  Nelson}]{pigolotti2013growth}
Pigolotti, S., Benzi, R., Perlekar, P., Jensen, M.~H., Toschi, F., Nelson, D.,
  2013. Growth, competition and cooperation in spatial population genetics.
  Theoretical population biology 84, 72--86.

\bibitem[{Purcell(1977)}]{purcell1977life}
Purcell, E.~M., 1977. Life at low reynolds number. Am. J. Phys 45~(1), 3--11.

\bibitem[{Visser et~al.(2009)Visser, Mariani, and
  Pigolotti}]{visser2009swimming}
Visser, A.~W., Mariani, P., Pigolotti, S., 2009. Swimming in turbulence:
  zooplankton fitness in terms of foraging efficiency and predation risk.
  Journal of Plankton Research 31~(2), 121--133.

\bibitem[{Waddell et~al.(2010)Waddell, Sander, and
  Doering}]{waddell2010demographic}
Waddell, J.~N., Sander, L.~M., Doering, C.~R., 2010. Demographic stochasticity
  versus spatial variation in the competition between fast and slow dispersers.
  Theoretical population biology 77~(4), 279--286.

\bibitem[{Whitlock(2003)}]{whitlock2003fixation}
Whitlock, M.~C., 2003. Fixation probability and time in subdivided populations.
  Genetics 164~(2), 767--779.

\end{thebibliography}







\end{document}